\documentclass[preprint2,tighten]{aastex6}
\pdfoutput=1 
\usepackage{amsmath,amstext}
\usepackage[T1]{fontenc}
\usepackage{apjfonts} 
\usepackage[figure,figure*]{hypcap}
\usepackage{color}
\usepackage{booktabs}
\usepackage{longtable}
\usepackage{multirow}

\newcommand{\unit}[1]{\ensuremath{\, \mathrm{#1}}}

\usepackage{hyperref}
\usepackage{breakurl}
\shorttitle{Diffuser-assisted follow-up of K2-28b and K2-100b}
\shortauthors{Stefansson \& Li et al. 2018}

\begin{document}
\title{Diffuser-assisted Photometric Follow-up Observations of the Neptune-sized planets K2-28b and K2-100b}
\author{Gudmundur Stefansson\altaffilmark{1,2,3}$^{,\dagger}$}
\author{Yiting Li\altaffilmark{1,4}$^{,\dagger}$}
\author{Suvrath Mahadevan\altaffilmark{1,2}}
\author{John Wisniewski\altaffilmark{5}}
\author{Leslie Hebb\altaffilmark{6}}
\author{Brett Morris\altaffilmark{7}}
\author{Joseph Huehnerhoff\altaffilmark{7,8,9}}
\author{Suzanne Hawley\altaffilmark{7}}

\email{gudmundur@psu.edu}
\thanks{$^{\dagger}$These authors contributed equally to this work.}
\altaffiltext{1}{Department of Astronomy \& Astrophysics, The Pennsylvania State University, 525 Davey Lab, University Park, PA 16802, USA}
\altaffiltext{2}{Center for Exoplanets \& Habitable Worlds, University Park, PA 16802, USA}
\altaffiltext{3}{NASA Earth and Space Science Fellow}
\altaffiltext{4}{Department of Physics, University of California, Santa Barbara, CA 93106, USA}
\altaffiltext{5}{Homer L. Dodge Department of Physics and Astronomy, University of Oklahoma, 440 W. Brooks Street, Norman, OK 73019, USA}
\altaffiltext{6}{Department of Physics, Hobart and William Smith Colleges, 300 Pulteney Street, Geneva, NY 14456, USA}
\altaffiltext{7}{Department of Astronomy, Box 351580, University of Washington, Seattle, WA 98195, USA}
\altaffiltext{8}{Apache Point Observatory, 2001 Apache Point Road, Sunspot, New Mexico, NM 88349, USA}
\altaffiltext{9}{Hindsight Imaging, Inc., 233 Harvard St. Suite 316, Brookline, MA 02446, USA}

\begin{abstract}
We present precision transit observations of the Neptune-sized planets K2-28b and K2-100b, using the Engineered Diffuser on the ARCTIC imager on the ARC 3.5m Telescope at Apache Point Observatory. K2-28b is a $R_{p} = 2.56 R_\oplus$ mini-Neptune transiting a bright (J=11.7) metal-rich M4 dwarf, offering compelling prospects for future atmospheric characterization. K2-100b is a $R_{p} = 3.45 R_\oplus$ Neptune in the Praesepe Cluster and is one of few planets known in a cluster transiting a host star bright enough ($V=10.5$) for precision radial velocity observations. Using the precision photometric capabilities of the diffuser/ARCTIC system, allows us to achieve a precision of $105^{+87}_{-37}$ppm, and $38^{+21}_{-11}$ppm in 30 minute bins for K2-28b, and K2-100b, respectively. Our joint-fits to the \textit{K2} and ground-based light-curves give an order of magnitude improvement in the orbital ephemeris for both planets, yielding a timing precision of $2 \unit{min}$ in the JWST era. Although we show that the currently available broad-band measurements of K2-28b's radius are currently too imprecise to place useful constraints on K2-28b's atmosphere, we demonstrate that JWST/NIRISS will be able to discern between a cloudy/clear atmosphere in a modest number of transit observations. Our light-curve of K2-100b marks the first transit follow-up observation of this challenging-to-observe transit, where we obtain a transit depth of $819 \pm 50 \unit{ppm}$ in the SDSS $i^\prime$ band. We conclude that diffuser-assisted photometry can play an important role in the TESS era to perform timely and precise follow-up of the expected bounty of TESS planet candidates.
\end{abstract}

\section{Introduction}
The original \textit{Kepler} mission \citep{borucki2010} discovered a wealth of exoplanets, yielding illuminating insights into the occurrence rates of exoplanets of different sizes \citep{dressing2015,petigura2013,burke2015,hsu2018}, the properties of multi-planet systems \citep{lissauer2011,fabrycky2014} and exoplanet mass-radius relations \citep{weiss2014,wolfgang2016,chen2017}. However, due to the faintness of the average \textit{Kepler} planet-host, most exoplanets discovered with \textit{Kepler} are not readily characterizable with facilities from the ground. There is a pressing need to detect such planets, as they can give us further insights into the exoplanet mass-radius relation \citep{weiss2014,wolfgang2016,chen2017}, further refine the structure of the exoplanet radius gap \citep{fulton2017}, and are the most favorable for further study with atmospheric transmission spectroscopy \citep[e.g.,][]{kempton2018}.

Since 2014, the re-purposed \textit{Kepler} mission, \textit{K2} \citep{Howell2014}, has shifted \textit{Kepler's} gaze to observe a larger fraction of planet-hosts that are more readily characterizable than from the original mission---namely to observe a larger fraction of brighter stars and a larger fraction of M-dwarfs. Since then, \textit{K2} has revealed a wealth of small planetary systems, including the sub-Neptunes HD106315c \citep{crossfield2017,rodriguez2017,lendl2017} and HD-3167c \citep{vanderburg2016}, given more insight into planets in clusters of different ages \cite[e.g.,][]{mann2016,mann2017k2_100}, discovered more ultra-short period planets \cite[e.g.,][]{malavolta2018,dai2017,adams2017} and warm Neptunes and Jupiters \citep[e.g.,][]{sphorer2017,yu2018}.

Further extending \textit{Kepler's} and \textit{K2's} legacy, the primary goal of the Transiting Exoplanet Survey Satellite (TESS) \citep{Ricker2014} is to find small planets that are most amenable for follow-up observations from the ground. One of the mission requirements of TESS is to measure the masses of 50 exoplanets with radii less than $4R_\oplus$. To do so, TESS will survey the nearest and brightest stars across the night sky for transiting exoplanets, observing most of the sky for at least 26 days. The expected yield from TESS has been studied by many groups \cite[e.g.,][]{sullivan2015,barclay2018,ballard2018}, and all groups predict that TESS will discover a multitude of exoplanets orbiting bright stars: hundreds of planets larger than Neptune and dozens of Earth-sized planets.

While TESS will help transform our understanding of the diversity of exoplanet systems around the closest stars, ground-based follow-up observations will play a key role in maximizing the scientific yield from TESS. In particular, precision transit follow-up photometry will be important in a number of ways.

First, precision photometry can be used to update orbital ephemerides, which will be essential for efficient scheduling for transit spectroscopy with JWST. For long-period planets, ephemerides refinement will be particularly important, as uncertain ephemerides are the main limitation to efficiently study these objects \citep{wang2015,dalba2016}. Additionally, planets with detectable Transit Timing Variations (TTVs) will allow us to gain key insights into the distribution of masses and composition of planets in such architectures \citep{agol2005,holman2005,mazeh2013}. Furthermore, obtaining precision photometry across different band-passes can be used as a planet confirmation tool \citep{tingley2014} and a method to obtain further information about the host star from the transit itself \citep{sandford2017}. Additionally, \cite{villanueva2016} predict that there will be 1218 single-transit events expected from TESS light curves, where 90\% of these events will have transit depths deeper than $1 \unit{mmag}$, making them amenable for photometric transit follow-up from current ground-based observatories. While space-based observatories such as \textit{Spitzer} have played an important role in conducting timely follow-up of exciting transiting systems \cite[see e.g.,][]{benneke2017,beichman2016}, the sheer number of expected planet candidates from the TESS sample produces a great need for high precision photometric instruments from the ground.

However, achieving high-precision photometry from the ground is difficult. Ground-based observations are subject to a number of limitations due to the day-night cycle, atmospheric effects, scintillation, transparency variations, differential extinction, seeing, and telescope-guiding effects \citep{mann2011,stefansson2017}. Beam-shaping diffusers \citep{stefansson2017} are emerging as an efficient and low-cost technology to provide stable high-precision photometry from the ground. These diffusers are nano-fabricated devices capable of molding the input starlight into a broad and stable top-hat image spread over many detector pixels. Spreading the light over many pixels in a stable manner allows us to minimize inter-pixel sensitivities on the detector and also allows us to increase exposure times and thus gather more photons. The additional increase in exposure time---and thus duty cycle---allows us to further better average over scintillation errors. Diffusers have now been used successfully at a number of telescopes \citep{stefansson2017,morris2018} and are actively being incorporated for high-precision photometry applications at a number of different observatories (Stefansson et al. 2018 in prep).

In this paper, we present diffuser-assisted photometric follow-up observations of two \textit{K2} planets: K2-28b and K2-100b first detected by \cite{hirano2016} and \cite{mann2017k2_100}, respectively. The transit observations in this paper were performed using Engineered diffuser \citep{stefansson2017} on the Astrophysical Research Council Telescope Imaging Camera (ARCTIC; \cite{huehnerhoff2016}) on the 3.5-meter Astrophysical Research Council (ARC) telescope at the Apache Point Observatory (APO). We reanalyze the \textit{K2} data of both planets using the Everest 2.0 \citep{luger2017} pipeline and combine the \textit{K2} data with our ground-based data to update the system parameters and orbital ephemerides.

K2-28b (EPIC 206318379) was confirmed by \cite{hirano2016} as a transiting Neptune-sized planet ($R_{p} = 2.56 R_\oplus$) using data from \textit{K2} Campaign 3, along with a combination of multi-band transit observations in the optical and NIR, low-resolution spectroscopy, and high-resolution adaptive-optics (AO) imaging. K2-28b is on a $P= 2.26$ day orbit around an M4 dwarf host star and has a similar radius and stellar irradiation to the well-studied GJ 1214b \citep{hirano2016}. Given the brightness of its host star in the near-infrared ($m_H=11.03$) and its relatively deep transit depth (6-7mmag), K2-28b is among one of few transiting planets around mid-to-late M dwarfs that facilitates precise follow-up observations, and has been discussed as a prime target for future atmospheric studies with the James Webb Space Telescope (JWST) \citep{hirano2016,chen2018}.

K2-100b (EPIC 211990866b) is a Neptune-size planet ($R_{p} = 3.45 R_\oplus$) on a $P=1.67 \unit{day}$ orbit around its $T_{\mathrm{eff}}=6120 \unit{K}$ host star in the 800 Myr Praesepe cluster, originally discovered by \cite{mann2017k2_100}. Given the brightness of its host star (V=10.65, J=9.46), K2-100b is one of few planets known in a cluster amenable for precise RV observations, leaving most other planets in clusters difficult to measure at-best \citep{mann2017k2_100}. Planets in clusters are compelling laboratories to test planet-formation and evolution models and are advantageous targets for precise planet-parameter estimations, as the properties of their host stars are generally better constrained than for their field-star counterparts \citep{mann2017k2_100,obermeier2016}.

This paper is structured as follows. Section 2 describes our diffuser-assisted transit follow-up observations of K2-28b and K2-100b with the ARC 3.5m telescope at APO. Section 3 describes our data reduction and calibration methods. In Section 4, we discuss our transit fitting of the \textit{K2} and ground-based light curves, and we present our main results in Section 5. In Section 6, we provide a further discussion of the importance of our ephemerides updates in the context of the JWST era, and discuss the possibility for future observations of these targets with radial velocity observations. We provide a short summary and conclusion of our main results in Section 7.

\section{Diffuser-Assisted Observations}

\subsection{Diffuser-assisted photometry}
We provide a brief discussion of diffusers here and refer the reader to a detailed discussion of diffusers and their application to precision photometry in \cite{stefansson2017}.

The diffuser available on ARCTIC is a customized Engineered Diffuser from RPC Photonics\footnote{\url{https://www.rpcphotonics.com/}}. Specifically, the ARCTIC diffuser is a 150mm-wide polymer-on-glass diffuser, produced using a precise laser-writing process. This laser process uses a modulated Ultra-Violet (UV) laser on a precision XY stage to write a structured pattern on a photoresist master, whose micro-structures are engineered to deliver a given diffusion angle and an a desired output PSF. The ARCTIC diffuser is engineered to produce a diffusion angle of $0.34\unit{^\circ}$, and delivers a fixed top-hat 9$\arcsec$ Full-Width-At-Half-Maximum (FWHM) PSF across the full field-of-view (FOV) on the ARCTIC detector. Being relatively inexpensive to make once a design and a master have been fabricated, similar Engineered Diffusers are available off the shelf with a wide range of opening angles, that can be relatively easily be adapted for use on telescopes large and small for precision photometry applications. As such, diffusers are now being tested and incorporated at a number of different telescopes and observatories for precision photometry---especially with TESS follow-up in mind, which will be further discussed in Stefansson et al. 2018b in prep.

\subsection{Observations}
We observed K2-28b on 2017 June 27 from 08:00-11:00 UT using the 3.5-meter ARC telescope at APO, using the SDSS i$^\prime$ filter and the Engineered Diffuser. K2-28 is a metal-rich M4-dwarf that has a V magnitude of 16.06, and SDSS $i^\prime$ magnitude of 13.9 \citep{hirano2016}. The target rose from airmass 2.19 at the start of the observations, to airmass 1.32 at the end of the observations. The night was nearly photometric, with seeing at $1\arcsec$ at the beginning of the night. Due to the faintness of the target, we used ARCTIC in 4$\times$4 binning mode. To minimize readout time and optimize the observing efficiency we read out the ARCTIC detector in quad-amplifier and fast-readout mode, yielding a $2.7 \unit{s}$ readout time. We set the exposure time to $30 \unit{s}$ to reach $\sim$5,000 peak counts per pixel for K2-28, resulting in a total observing cadence of 32.7s. We obtained 51 bias and 21 dark frames, along with 51 dome flats to correct the data following standard procedures outlined in \cite{stefansson2017}.

We observed K2-100b on 2018 March 16 from 03:00-06:00 UT using the 3.5m telescope at APO using the SDSS $i^\prime$ filter, and the Engineered Diffuser. K2-100 has an r$^\prime$ magnitude of 10.373, and SDSS i$^\prime$ magnitude of 10.22 \citep{zacharias2012}. The target rose during the night, starting at airmass 1.17, peaking at 1.02 at the meridian, and ending at airmass 1.17 at the end of the observations. We initially set the exposure time to 15s, but changed it to 14s after 15 minutes of observations to keep the peak counts in the target within the linear regime of the detector ($<55,000$ counts per pixel). To maximize the observing efficiency similar to the K2-28b observations, we obtained the K2-100 exposures in ARCTIC quad-amplifier fast-readout 4$\times$4 binning mode, resulting in a mean observing cadence of 15.7s. The night was non-photometric, with a seeing of 3$\arcsec$ at the start. We obtained a set of 25 biases and 101 dome flats for standard calibrations following \cite{stefansson2017}.

\section{Data Reduction}
We used the AstroImageJ software \citep{collins2017} to reduce our photometric datasets, including the creation of master calibration frames (bias, darks and flats), along with using it to perform the photometric extraction. Before running the photometry through AstroImageJ, we tried cleaning both of our ground-based datasets for cosmic rays using the \texttt{astroscrappy} package which uses the Laplacian-edge cosmic-ray rejection algorithm presented in \cite{Dokkum2011}, and found that it improved the photometry for K2-28b, but not K2-100b. Therefore, we used the cosmic-ray cleaned data for K2-28b, and used the uncleaned data for K2-100b. 

For our K2-28b observations, we performed differential aperture photometry in AstroImageJ. Five reference stars were used, two of which are of comparable brightness to the target star and three of which are 30-50\% of the brightness of the target star. We found that the aperture setting that yielded the smallest residuals was $18$, $32$, and $48 \unit{pixels}$ for the aperture radius, inner annuli and outer annuli, respectively. After removing an additional 10 significant ($>3\sigma$) outliers present in the data, we obtained an unbinned precision of $3830 \unit{ppm}$.

For our K2-100b observations, five reference stars were used in the photometric reduction process. Two of the reference stars were relatively faint (10-30\% of the brightness of the target star) while the other three were as bright as the target star. We tried a few different photometric apertures, and observed that the aperture setting that gave the smallest residuals was $20$, $32$, and $50 \unit{pixels}$ for the aperture radius, and the inner and outer annuli respectively. The raw data showed a slightly sloped light curve, and appeared to be affected by a few clouds passing during the observations. We checked for correlations with sky background, airmass, time, and X and Y pixel centroid coordinates and found that detrending with time would take out the slope. We simultaneously fit for this slope in our MCMC fitting, as further discussed below.

\section{Transit Fitting}
We modeled the transits in a Markov-Chain Monte Carlo (MCMC) framework, using the affine-invariant \textit{emcee} Python package \citep{dfm2013} to perform the MCMC sampling. To calculate the transit model, we used the \textit{batman} package \citep{kreidberg2015}, which uses the \cite{mandel2002} transit formalism.

We adopt the usual $\chi^2$ likelihood function to find the maximum likelihood solution. Before initializing the Markov Chains, we used a differential evolution optimizer \citep{pyde} to find the global maximum likelihood solution, and then we initialize 100 \texttt{emcee} walkers in the vicinity of this solution to sample the parameter space for 10,000 steps each. Depending on the fit, we removed the first 2000 steps as burn-in. We used the Gelman-Rubin test to check for convergence, considering chains with a Gelman-Rubin statistic \cite[e.g.,][]{ford2006} within $5\%$ of unity as well mixed. The resultant posteriors were visually smooth and unimodal. 

For each planet, we performed 3 fits: a fit of the K2 data, a fit of the ground-based data, and finally a joint fit of both datasets fitted simultaneously. For all fits, we followed the suggestion in \cite{Eastman2013}, using the following parameters as MCMC jump parameters: the logarithm of the orbital period $\log(P)$, the transit midpoint $T_C$, the radius ratio $R_p/R_*$, the cosine of the inclination $\cos(i)$, and the logarithm of the scaled semi-major axis, $\log(a/R_*)$, and finally a parameter describing the transit baseline flux. Following our previous work in \cite{stefansson2017}, for all of the fits, we assumed an eccentricity of 0 and fixed quadratic limb-darkening parameters. We calculated the expected limb-darkening parameters assuming a quadratic limb-darkening law from \cite{claret2011}, using the EXOFAST limb-darkening calculator web-applet\footnote{\url{http://astroutils.astronomy.ohio-state.edu/exofast/limbdark.shtml}}. To calculate the limb darkening parameters, we used the host star $\log g$, $T_{\mathrm{eff}}$, and $\mathrm{[Fe/H]}$ values listed in Table \ref{tab:stellarpriors} for the $Kepler$ and/or SDSS $i^\prime$ bands, depending on the observations being analyzed. Our priors for our MCMC fits are summarized in Table \ref{tab:priors}, where we opted to impose broad uniform priors on most jump parameters to give the walkers freedom to explore a broad range of parameter space, and to impose a minimal prior bias on our results. 

We discuss our individual fits in further detail below. All of the data analysis performed in this paper is captured in Jupyter notebooks accessible on GitHub\footnote{\url{https://github.com/gummiks/Diffuser-Assisted-K2-Followup}}.

\begin{deluxetable*}{llcc}
\tablecaption{Stellar parameters for K2-28 and K2-100 used in this work. The stellar parameters are adopted from \cite{hirano2016} and \cite{mann2017k2_100} for K2-28 and K2-100, respectively. The limb-darkening parameters were calculated using the EXOFAST web-applet for the different band-passes, using the stellar parameters in this table.\label{tab:stellarpriors}}
\tablehead{\colhead{Parameter}         						      &  \colhead{Description}                  & \colhead{K2-28b}           & \colhead{K2-100b}           }
\startdata
$M_* (M_\odot)$                                                   &  Stellar mass                           & $0.257\pm0.048$              &  $1.18\pm 0.09$           \\
$R_* (R_\odot)$                                                   &  Stellar radius                         & $0.288\pm0.028$              &  $1.19\pm0.05$            \\
$\rho_* (\rho_\odot)$                                             &  Stellar density                        & $15.2\pm2.4$                 &  $0.99\pm 0.14$           \\
$T_{\mathrm{eff}}$ (K)                                            &  Stellar effective temperature          & $3214\pm60$                  &  $6120\pm 90$             \\
$\mathrm{[Fe/H]}$                                                 &  Stellar metallicity                    & $0.26\pm0.05$                &  $0.14\pm0.04$            \\
$\log(g)$                                                         &  Stellar gravity                        & $4.93\pm 0.04$               &  $4.360\pm0.03$           \\
$u_\mathrm{1,K2}$                                                 &  Linear limb-darkening coefficient      & 0.4266                       & 0.3490                    \\
$u_\mathrm{2,K2}$                                                 &  Quadratic limb-darkening coefficient   & 0.3076                       & 0.2937                    \\
$u_\mathrm{1,ground}$                                             &  Linear limb-darkening coefficient      & 0.3402                       & 0.2622                    \\
$u_\mathrm{2,ground}$                                             &  Quadratic limb-darkening coefficient   & 0.3155                       & 0.3017                    \\
\enddata
\end{deluxetable*}

\begin{deluxetable*}{llccc}
\tablecaption{Summary of priors for K2-28b and K2-100b. Priors on stellar parameters for K2-28 are adopted from \cite{hirano2016}, and from \cite{mann2017k2_100} for K2-100. $\mathcal{N}(m,\sigma)$ denotes a normal prior with a mean $m$, and standard deviation $\sigma$; $\mathcal{U}(a,b)$ denotes a uniform prior with a start value $a$ and end value $b$. \label{tab:priors}}
\tablehead{\colhead{Parameter}         						      &  \colhead{Description}                  & \colhead{K2-28b}             							& \colhead{K2-100b}      } 
\startdata	
\hline
\multicolumn{4}{c}{\hspace{-0.3cm} \textit{K2}-only fit}           \\ %-----------------------------------------------------------------------------------------------------------------------------------------------------
\hline
$\log(P)$ (days)                                                  &  Orbital period                         & $\mathcal{U}(0.353,0.355)$                            & $\mathcal{U}(0.222,0.224)$            		 \\
$T_{\mathrm{C},K2}$                                               &  Transit Midpoint $(\mathrm{BJD_{TDB}})$& $\mathcal{U}(2456977.98,22456978.00)$                 & $\mathcal{U}(2457140.71,2457140.73)$           \\
$(R_p/R_*)_{\mathrm{K2}}$                                         &  Radius ratio                           & $\mathcal{U}(0,0.1)$                                  & $\mathcal{U}(0,0.1)$    						 \\
$\cos(i)$                                                         &  Transit inclination                    & $\mathcal{U}(0,0.2)$         							& $\mathcal{U}(0,0.2)$      					 \\
$\log(a/R_*)$                                                     &  Normalized orbital radius              & $\mathcal{U}(0.9,2.0)$       							& $\mathcal{U}(0.8,1.0)$    					 \\
$\sqrt{e} \cos(\omega) $                                          &  Eccentricity \& Argument of periastron & 0 (adopted)                 							& 0 (adopted)               					 \\
$\sqrt{e} \sin(\omega) $                                          &  Eccentricity \& Argument of periastron & 0 (adopted)                  							& 0 (adopted)               					 \\
fraw$_{\mathrm{K2}}$                                              &  Transit baseline for K2 data           & $\mathcal{U}(0.9,1.1)$                                & $\mathcal{U}(0.9,1.1)$  						 \\
$\sigma_{K2} $                                                    &  Average error for K2 data              & $\mathcal{U}(1\times10^{-5},5\times10^{-4})$          & $\mathcal{U}(1\times10^{-5},5\times10^{-4})$   \\
\hline
\multicolumn{4}{c}{\hspace{-0.3cm} Ground-based-only fit} \\ \hline %-----------------------------------------------------------------------------------------------------------------------------------------------------
$\log(P)$ (days)                                                  &  Orbital period                         & $\mathcal{N}(0.35419,0.00002)$          			    & $\mathcal{N}(0.223731,0.000012)$    			 \\
$T_{\mathrm{C,ground}}$                                           &  Transit Midpoint $(\mathrm{BJD_{TDB}})$& $\mathcal{U}(2457931.88,2457931.92)$           		& $\mathcal{U}(2457828.68,2457828.70)$           \\
$(R_p/R_*)_{\mathrm{ground}}$                                     &  Radius ratio                           & $\mathcal{U}(0,0.1)$         							& $\mathcal{U}(0,0.05)$    						 \\
$\cos(i)$                                                         &  Transit inclination                    & $\mathcal{U}(0,0.2)$         							& $\mathcal{U}(0,0.2)$      					 \\
$\log(a/R_*)$                                                     &  Normalized orbital radius              & $\mathcal{U}(0.9,2.0)$       							& $\mathcal{U}(0.8,1.0)$    					 \\
$\sqrt{e} \cos(\omega) $                                          &  Eccentricity \& Argument of periastron & 0 (adopted)                 							& 0 (adopted)               					 \\
$\sqrt{e} \sin(\omega) $                                          &  Eccentricity \& Argument of periastron & 0 (adopted)                  							& 0 (adopted)               					 \\
fraw$_{\mathrm{Ground}}$                                          &  Transit baseline for ground-based data & $\mathcal{U}(0.9,1.1)$       							& $\mathcal{U}(0.9,1.1)$   						 \\
$D_{\mathrm{Line}}$                                               &  Ground detrend parameter: line         & --           											& $\mathcal{U}(-0.1,0.1)$       				 \\
\hline
\multicolumn{4}{c}{\hspace{-0.3cm} Joint-fit}             \\ \hline %-----------------------------------------------------------------------------------------------------------------------------------------------------
$\log(P)$ (days)                                                  &  Orbital period                         & $\mathcal{U}(0.352,0.356)$                    		& $\mathcal{U}(0.222,0.224)$      		    	 \\
$T_{\mathrm{C,joint}}$                                            &  Transit Midpoint $(\mathrm{BJD_{TDB}})$& $\mathcal{U}(2457931.85,2457931.95)$          		& $\mathcal{U}(2457828.68,2457828.70)$           \\
$(R_p/R_*)_{\mathrm{ground}}$                                     &  Radius ratio                           & $\mathcal{U}(0,0.1)$         							& $\mathcal{U}(0,0.05)$   				    	 \\
$(R_p/R_*)_{\mathrm{K2}}$                                         &  Radius ratio                           & $\mathcal{U}(0,0.1)$         							& $\mathcal{U}(0,0.1)$    					     \\
$\cos(i)$                                                         &  Transit inclination                    & $\mathcal{U}(0,0.2)$         							& $\mathcal{U}(0,0.2)$      					 \\
$\log(a/R_*)$                                                     &  Normalized orbital radius              & $\mathcal{U}(0.9,2.0)$       							& $\mathcal{U}(0.8.1.0)$    					 \\
$\sqrt{e} \cos(\omega) $                                          &  Eccentricity \& Argument of periastron & 0 (adopted)                 							& 0 (adopted)               					 \\
$\sqrt{e} \sin(\omega) $                                          &  Eccentricity \& Argument of periastron & 0 (adopted)                  							& 0 (adopted)               					 \\
$\sigma_{K2} $                                                    &  Average error for K2 data              & $\mathcal{U}(1\times10^{-5},5\times10^{-4})$          & $\mathcal{U}(1\times10^{-5},5\times10^{-4})$   \\
fraw$_{\mathrm{Ground}}$                                          &  Transit baseline for ground-based data & $\mathcal{U}(0.9,1.1)$       							& $\mathcal{U}(0.9,1.1)$ 						 \\
fraw$_{\mathrm{K2}}$                                              &  Transit baseline for K2 data           & $\mathcal{U}(0.9,1.1)$      							& $\mathcal{U}(0.9,1.1)$  						 \\
$D_{\mathrm{Line}}$                                               &  Ground detrend parammeter: line        & --           											& $\mathcal{U}(-0.1,0.1)$       				 \\
\enddata
\end{deluxetable*}

\subsection{K2 Light Curve Analysis}
We detrended the K2 data using the Everest pipeline \citep{luger2016,luger2017}, which uses a combination of Gaussian-Processes (GP) and pixel-level decorrelation to detrend \textit{K2} light-curves. To independently recover the orbital period and epoch of first transit for both planets, we used a Box-Least-Squares (BLS) algorithm \citep{kovacs2002} implemented in Python and Fortran available on GitHub\footnote{\url{https://github.com/dfm/python-bls}}. Before running the BLS search algorithm, we flattened the K2 light curve using the best-fit Gaussian-Process model in Everest describing slowly-varying trends in the \textit{K2} light curve (e.g., due to stellar variability or slow instrumental changes due to temperature). To ensure that the GP model was not impacting or over-fitting the transits themselves, after finding the location of the transits using the BLS algorithm from this first flattening step, we recomputed the Everest GP systematic model---using the \texttt{.compute()} function in Everest---with all of the in-transit datapoints masked out. We then use this improved GP systematic model to flatten the \textit{K2} light curve for subsequent transit MCMC analysis. 

To further improve the period estimate on the transits, we performed all of our \textit{K2} fits on the unfolded \textit{K2} data. To reduce the data volume, we only analyzed the data obtained within a small window of $0.2\unit{days}$ surrounding the transit centers found by the BLS algorithm. Following \cite{kipping2010}, we over-sampled and binned our \textit{K2} transit model to match the 30 minute \textit{K2} cadence, using the \texttt{supersample\_factor = 30}, and \texttt{exp\_time = 0.02} keywords in \texttt{batman}. In addition to the jump parameters discussed above ($\log(P)$, $T_C$, $R_p/R_*$, $\cos(i)$, $\log(a/R_*$), and a transit baseline), we also included an independent fit parameter describing the average errorbar of the K2 light curve. We found that the average best-fit errorbar in the \textit{K2} data ranged was $50$ppm in 30 minute bins for K2-28b, and $40 \unit{ppm}$ for K2-100b.

\subsection{Ground-based Light Curve Analysis}
For our ground-based transit analysis we follow the steps outlined in \cite{stefansson2017}. To summarize that discussion, we calculate the total photometric errorbars using the following equation,
\begin{equation}
\sigma_{\mathrm{tot}} = \sqrt{ \sigma_{\textrm{rel flux}}^2 + \sigma_{\mathrm{scint}}^2},
\label{eq:stds}
\end{equation}
where $\sigma_{\textrm{rel flux}}$ is calculated using Equation 1 and 2 in \cite{stefansson2017}, and includes the photon, dark, readout and digitization noise. We calculate the scintillation term for a given star in units of relative flux as described in \cite{young1967} and \cite{dravins1998}, and further expanded by \cite{osborn2015}, with the following approximation:
\begin{equation}
\sigma_{s} = 0.135 D^{-\frac{2}{3}} \chi^{1.75} {\left(2t_{\mathrm{int}}\right)}^{-\frac{1}{2}} e^{\frac{-h}{h_0}} \sqrt{1 + 1/n_{\mathrm{E}}}
\label{eq:stds2}
\end{equation}
where $D$ is the diameter of the telescope in centimeters, $\chi$ is the airmass of the observation, $t_{int}$ is the exposure time in seconds, and $h$ is the altitude of the telescope in meters, and $h_0\simeq8000\unit{m}$ is the atmospheric scale height. The constant 0.135 factor in front has a unit of $\unit{cm^{2/3}s^{1/2}}$, to give the scintillation error in units of relative flux. Following the suggestion by \cite{osborn2015}, we note that we have multiplied this constant factor by 1.5 from the $0.09 \unit{cm^{2/3}s^{1/2}}$ value originally presented in \cite{young1967} and \cite{dravins1998}, to better reflect the median value of scintillation noise. Finally, the $\sqrt{1 + 1/n_{\mathrm{E}}}$ reference star term is derived in \cite{kornilov2012}, and describes the number of uncorrelated reference stars $n_\mathrm{E}$. For our observations, we assumed that all our reference stars were uncorrelated.

To remove any systematics in the light-curves, we explored using different simultaneous detrending parameters, including airmass, X-and-Y centroid coordinates, and/or a straight line. From our exploration we determined that adding detrending parameters to the K2-28b data did not yield a significant improvement to our fits. For the K2-100b data, we observed that a simple linear slope (detrending with the time coordinate) was sufficient to take out a small correlated slope we observed in the raw photometry.

\subsection{Joint K2 and Ground-based Light Curve Analysis}
After fitting both the K2 data and the ground based separately, we performed a joint global fit including both the ground based and \textit{K2} data, following the data preparation steps discussed above. For this fit, we assume that the planets follow a strict periodic orbit with no transit timing variations.

We perform the joint fits for K2-28b and K2-100b with 9 and 10 independent parameters, respectively. Following \cite{chen2018}, we allowed $R_p/R_*$ to vary separately in the ground-based and the \textit{K2} data, while assuming common values for $\log(P)$, $T_C$, $\log(a/R_*)$, and $\cos(i)$. Similarly for the \textit{K2}-only fits, we include one parameter describing the mean error in the \textit{K2} data. Additionally, we include two independent parameters describing the transit baselines for the \textit{K2} and ground-based data. Finally, for the K2-100b data, we included an additional detrending parameter describing the linear slope in the data for a total of 10 independent parameters. The priors we used for these parameters are summarized in Table \ref{tab:priors}.

\begin{figure*}[t]
	\begin{center}
		\includegraphics[width=0.9\textwidth]{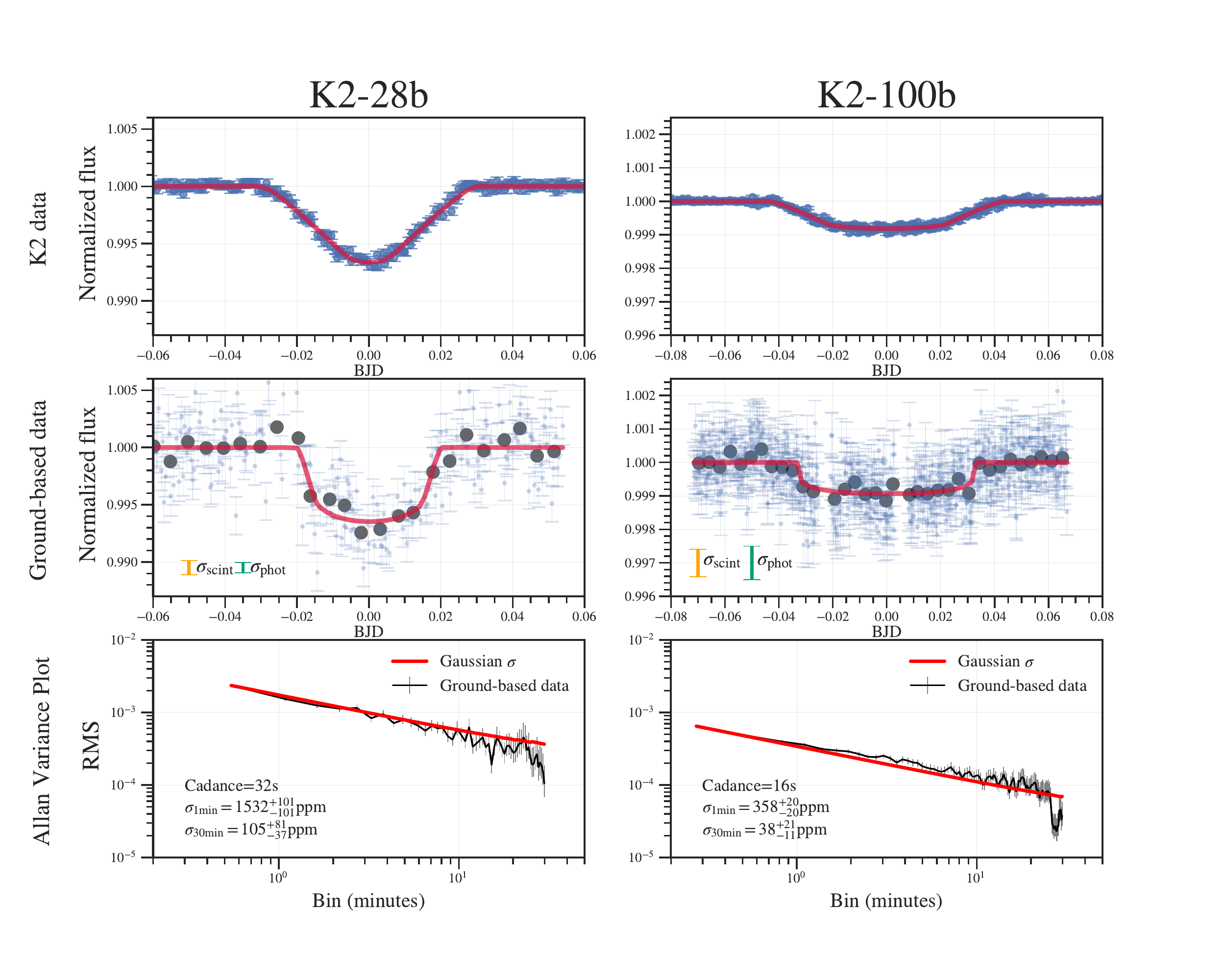}
	\end{center}
	\caption{Transits of a) K2-28b and b) K2-100b. Top panels: Best-fit phase-folded transits of the \textit{K2} data analyzed in this work. Middle panels: Diffuser-assisted ground-based transits as observed with the diffuser on the ARC 3.5m Telescope at APO. The unbinned data are shown in blue with a cadence of 32s and 16s for K2-28b and K2-100b, respectively. The gray points are 5 minute bin points. Additionally shown are the mean errorbars from photon, dark, and read noise (green errorbar; $\sigma_{\mathrm{phot}}$) and the errorbar due to scintillation (orange errorbar; $\sigma_{\mathrm{scint}}$). The gaps in the K2-100b were due to clouds. Bottom panels: Photometric precision as a function of bin size in minutes.} 
	\label{fig:transits}
\end{figure*}

\section{Results}
Figure \ref{fig:transits} shows the transits studied in this work. The top panels show the the best-fit phase-folded \textit{K2} data, and the middle panel shows our ground-based transits. The ground-based transits are shown with errorbars including photon, read, dark and scintillation noise, as calculated by equation \ref{eq:stds}. Following \cite{stefansson2017}, in the middle panel, we compare the mean photometric errorbar from photon, read and dark noise ($\sigma_{\mathrm{phot}}$; shown in green) to the errorbar due to scintillation only ($\sigma_{\mathrm{scint}}$; shown in orange). We estimated the mean scintillation errorbars using the mean airmasses of 1.58 and 1.06, for our K2-28b and K2-100b observations, respectively. For both observations, we see that the scintillation errors similar in magnitude to the total errorbar due to photon, read and dark noise.

The bottom panels in Figure \ref{fig:transits} show the standard deviation of the best-fit residuals as a function bin size for our ground-based observations, calculated using the \texttt{MC3} code \citep{cubillos2017} which produces errorbars assuming the RMS scatter follows an inverse-gamma distribution for the highest bin sizes. Additionally shown in red is the Gaussian expected precision, assuming white noise behavior with increasing binning sizes. For K2-28b, we achieve a precision of $\sigma_{\mathrm{30min}}=105_{-81}^{+37}\unit{ppm}$, and a precision of $\sigma_{\mathrm{30min}}=38_{-11}^{+21}\unit{ppm}$, for K2-100b. For our K2-100b observations, we see statistical fluctuations at the highest binning times, i.e., where the number of bins are the smallest, dipping below the Gaussian expected precision. The expected Gaussian precision for K2-100b is $80 \unit{ppm}$ in 30 minutes bins, which is a more conservative estimate of the precision at that binning level.

\section{Discussion}
The ground-based diffuser-assisted observations presented here allow us to improve the orbital parameters for K2-28b and K2-100b in two key areas. First, our observations jointly modeled with the \textit{K2} data, provide a significant improvement to the orbital ephemerides  from the \textit{K2} data alone. For K2-28b, we closely compare our orbital ephemerides update to the recent work by \cite{chen2018}. Second, our improved observing cadence from the ground allows us to resolve the transit shape better than in the under-sampled V-shaped \textit{K2} transits, giving us better constraints on the orbital inclination $i$, and the orbital semi-major axis $a/R_*$ for both planets. Additionally, for K2-28b, we take a closer look at the available broad-band transit depth values for K2-28b, and look for any emerging patterns in its radius as a function of wavelength. We finally end our discussion about the feasibility for future follow-up observations with radial velocity observations.

\subsection{Improved Orbital Ephemerides}
\begin{figure*}[t]
	\begin{center}
		\includegraphics[width=0.95\textwidth]{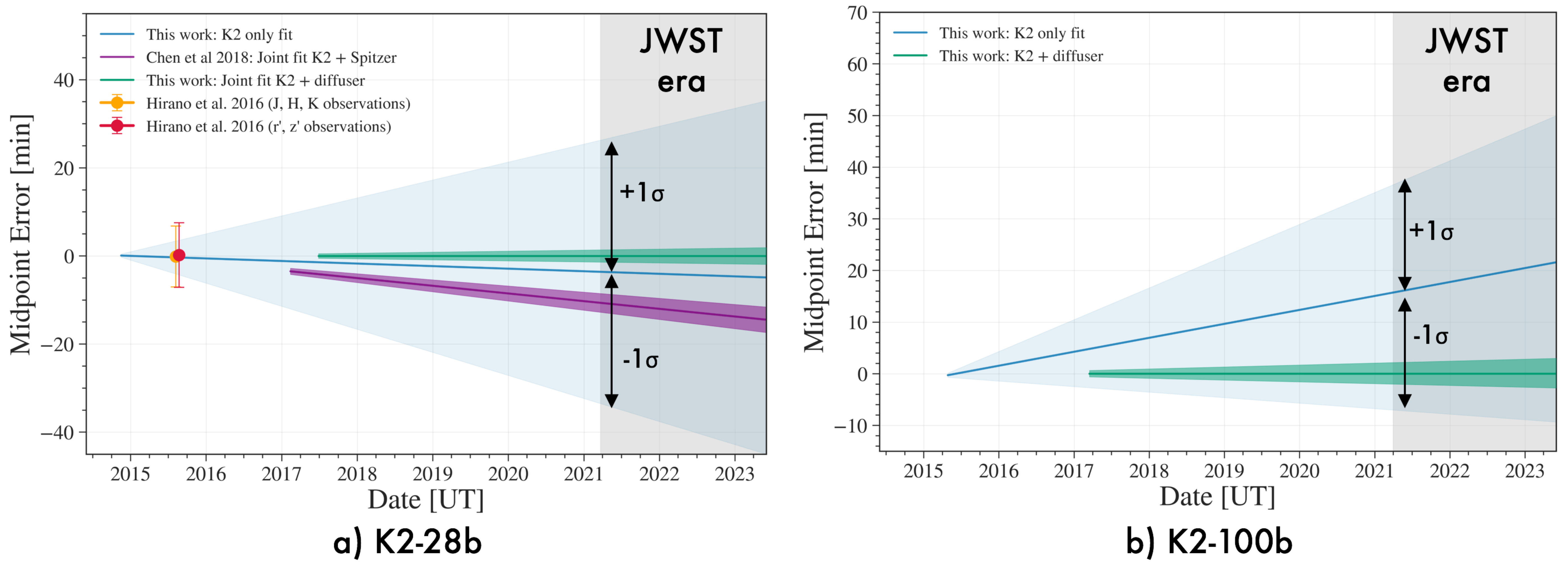}
	\end{center}
	\caption{Updated ephemerides for a) K2-28b and b) K2-100b. The shaded blue regions show the ephemerides derived from the \textit{K2}-only fits in this work, and the green-shaded regions show the ephemerides derived from our joint \textit{K2} and ground-based fits. A nominal beginning of the JWST-era is shown in the gray-shaded area, assuming a launch date of March 30th, 2021. Additionally shown for K2-28b (left panel) are midpoints derived from \cite{hirano2016} (orange and red points), along with a comparison to the \cite{chen2018} \textit{K2}+\textit{Spitzer} ephemeris.}
	\label{fig:ephemupdate}
\end{figure*}

 Our ground-based observations extend the observing baseline from the 80-day \textit{K2} observing baseline to a baseline of over 2-years for both planets, directly resulting in an improved measurement of the orbital ephemeris from our joint fits. Figure \ref{fig:ephemupdate} compares our ephemerides from our \textit{K2}-only analysis (in blue) with our updated joint-fit ephemerides (in green). We see that our updated transit ephemerides are in both cases consistent with the ephemeris derived from the \textit{K2} data alone, within the 1$\sigma$ shaded region. Additionally, Figure \ref{fig:ephemupdate} demonstrates that at the start of the nominal JWST era in March 2021, the 1-$\sigma$ ephemerides errors improved by an order of magnitude by adding our ground-based observations to our \textit{K2}-only analysis---from $25 \unit{min}$ down to $\sim$$1.5 \unit{min}$ for both planets---allowing for efficient scheduling of future JWST observations.

For K2-28b, we further compare our updated ephemeris with the ephemeris in \cite{chen2018} (shown in purple in Figure \ref{fig:ephemupdate}), and the transits observed in \cite{hirano2016} (orange, and red points in Figure \ref{fig:ephemupdate}). From our joint \textit{K2}-and ground-based diffuser-assisted observations, we derive an ephemeris of,
\begin{equation}
\begin{aligned}
P_{\mathrm{joint}}   &= 2.2604455 \pm 0.0000010, \\
T_{\mathrm{C,joint}} &= 2457931.89780_{-0.00037}^{+0.00036}.
\end{aligned}
\end{equation}
This ephemeris differs slightly from the ephemeris presented in \cite{chen2018},
\begin{equation}
\begin{aligned}
P_{\mathrm{joint}}   &= 2.2604380 \pm 0.0000015, \\
T_{\mathrm{C,joint}} &= 2457796.26865_{-0.00049}^{+0.00048},
\end{aligned}
\end{equation}
as is evident from the non-overlapping green and purple regions in Figure \ref{fig:ephemupdate}. We quantify the disagreement to be at the 3$\sigma$ level at the time of our observations on June 26th 2017 UT, and at the 10 minute level at the start of the JWST era in 2021. This discrepancy could be caused by two main sources. First, systematics could be present in both or either of the light-curves slightly offsetting the transit midpoints, impacting the derived planet period. In particular, \cite{chen2018} report a modest $<1.4\sigma$ dependence between the length of the out-of-transit baseline they used to detrend the transit, and the value of their transit center $T_C$, which could bring our transit midpoints to within $2 \sigma$ agreement in some cases. Second, this difference could be caused by potential transit timing variations, which could be indicative of a second planet orbiting in the system. We reserve transit timing variation analysis for the K2-28b system for future efforts. Although modest, this discrepancy highlights the importance of repeated follow-up observations to reduce systematic uncertainties in transit ephemerides. 

\subsection{Improved Orbital Parameters}
Our faster observing cadence obtained from the ground than the 30 minute \textit{K2} cadence allows us to resolve the transit shape better, yielding tighter constraints on the inclination $i$ and the scaled semi-major axis $a/R_*$ for both transits. From Table \ref{tab:k228bmcmc} and Table \ref{tab:k2100bmcmc}, we see that our ground-based values agree to within $1 \sigma$ to our \textit{K2}-only analysis values, providing further evidence that these transits are not due to an astrophysical false positive (e.g., a blended stellar eclipsing binary).

For K2-28b, we prefer a slightly larger radius of $R_p = 2.56^{+0.27}_{-0.26} R_\oplus$ in the \textit{K2} band, than the radius from \cite{hirano2016} of $R_p = 2.32 \pm 0.24 R_\oplus$, and the radius from \cite{chen2018} of $R_p = 2.30^{+0.26}_{-0.28} R_\oplus$, but all values agree within $1\sigma$. We further note that our joint fit $R_p = 2.48 \pm 0.26 R_\oplus$ radius in the SDSS i$^\prime$ band, agrees well with the \textit{Spitzer} radius presented in \cite{chen2018} of $R_p = 2.45 \pm 0.28 R_\oplus$. For K2-100b, our radius of $R_p = 3.45^{+0.16}_{-0.15} R_\oplus$ in the \textit{K2} band, and the moderately larger $R_p = 3.71^{+0.20}_{-0.19} R_\oplus$ in the SDSS i$^{\prime}$ band, agrees well with the radius from \cite{mann2017k2_100} of $R_p = 3.5 \pm 0.2 R_\oplus$. With respect to K2-100b's transit depth, our observations yield a transit depth of $819 \pm 50 \unit{ppm}$ in the SDSS $i^\prime$ band, consistent with our \textit{K2}-only transit depth of $689^{+6}_{-5}\unit{ppm}$ at the $2-3\sigma$ level. We further discuss the radius of K2-28b as a function of wavelength in further detail in the following subsection below.

%-------------------------------------------------------------
%-------------------------------------------------------------
\subsection{K2-28b: broad-band radii as a function of wavelength}
\begin{figure*}[t]
	\begin{center}
		\includegraphics[width=0.80\textwidth]{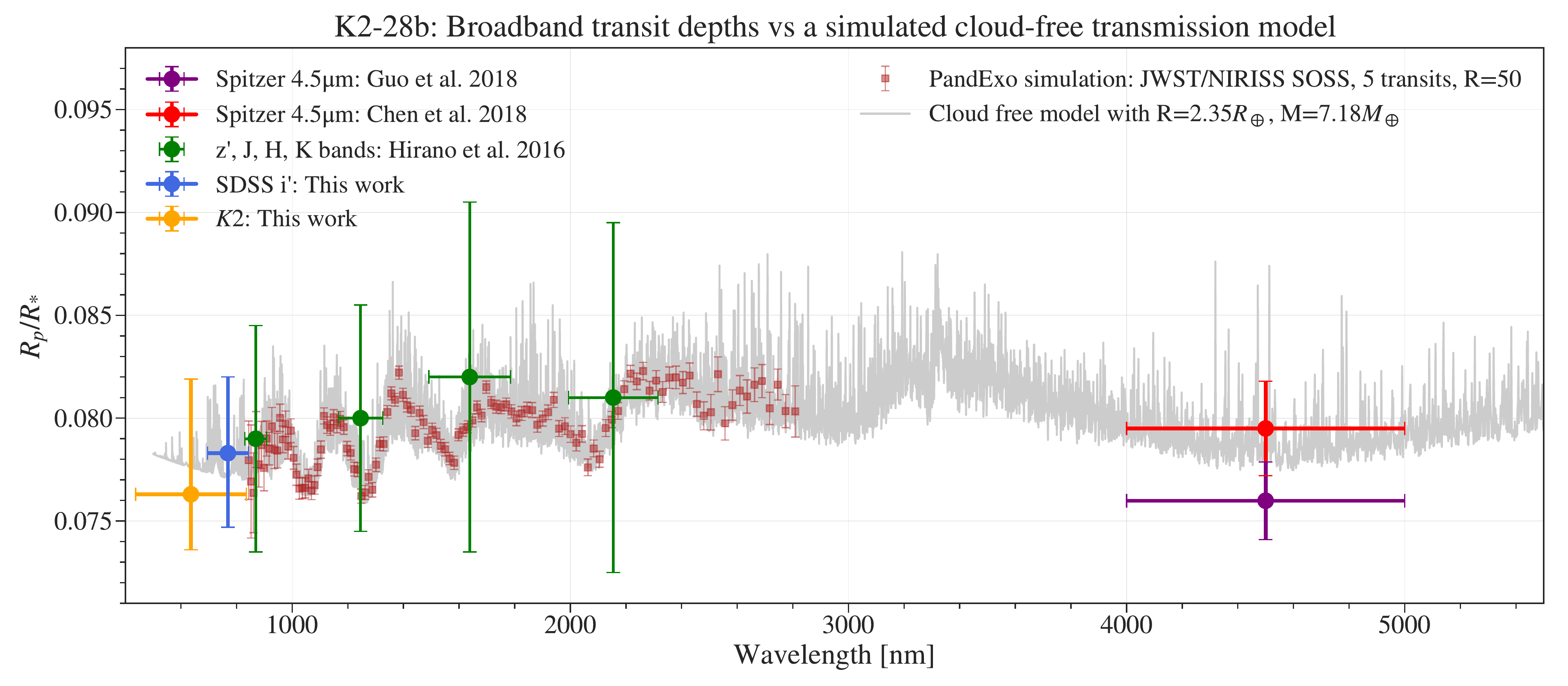}
	\end{center}
	\caption{Comparison of $R_p/R_*$ values of K2-28b in different bands from this work, showing diffuser assisted observations (blue point) \textit{K2} data (orange point), ground-based observations from \cite{hirano2016} (green points), and \textit{Spitzer} observations from \cite{chen2018} and \cite{guo2018} (red and purple points). Additionally overlaid is a model cloud-free transmission-spectrum of K2-28b calculated using \texttt{PandExo}, assuming a predicted mass of $7.18 \unit{M_\oplus}$. We see that the current broad-band observations are not precise enough to draw any conclusions regarding the atmospheric structure of K2-28b. However, if K2-28b has a clear atmosphere, we do show that JWST/SOSS will be able to resolve its atmospheric structures in a modest number of 5 transits.}
\label{fig:transmission}
\end{figure*}

To study the emerging picture of K2-28b's transmission spectrum from the growing number of available ground and space based radius measurements, in Figure \ref{fig:transmission} we compare the measured $R_p/R_*$ from this work from our \textit{K2} and ground-based analyses, along with the ground-based $z^\prime$, $J$, $H$ and $K_S$ values from \cite{hirano2016}, and the \textit{Spitzer} $4.5 \unit{\mu m}$ $R_p/R_*$ values presented in \cite{chen2018}, and an independent analysis of the same \textit{Spitzer} transit by \cite{guo2018} (see numeric values in Table \ref{tab:k228bradius}). We note that the SDSS $r^\prime$ band radius measurement presented in \cite{hirano2016} of $R_p/R_* = 0.063_{-0.011}^{+0.009}$ is discrepant to $\sim$2$\sigma$ with the other values in Table \ref{tab:k228bradius}, potentially due to systematics present in the partial transit observations. For clarity, we decided to omit this point from our overview in Figure \ref{fig:transmission}. Furthermore, we note that \cite{hirano2016} present two separate analyses of their ground-based observations: A) an analysis that imposed an airmass cutoff of <2.2 for the partial-transit SDSS $r^\prime$ and $z^\prime$ observations, and where they used time-varying photometric apertures to extract the $J$, $H$, and $K_S$ observations, and B) an analysis without any airmass cutoffs, and where they used fixed photometric apertures for all observations. Here we specifically chose the former analysis, which we deemed more robust, and less susceptible to systematics. Lastly, from Figure \ref{fig:transmission} we see a difference in the \textit{Spitzer} analyses to within $1\sigma$ presented in \cite{guo2018} and \cite{chen2018}, which will be further discussed below.

\begin{deluxetable}{lccr}
\tabletypesize{\scriptsize}
\tablecaption{$R_P/R_*$ measurements in different bands for K2-28b. The SDSS $r^\prime$ radius measurement from \cite{hirano2016} is discrepant to $\sim$2$\sigma$, and we did not include it in Figure \ref{fig:transmission} for clarity.\label{tab:k228bradius}}
\tablehead{
\colhead{Band} & \colhead{Band Center} & \colhead{$R_p/R_*$} & \colhead{Reference}\\
\colhead{} & \colhead{(nm)} & \colhead{} & \colhead{} 
}
\startdata
SDSS $r^\prime$                     & 630  & $0.056^{+0.009}_{-0.010}$    & \cite{hirano2016} \\
$Kepler/K2$ 			            & 636  & $0.0763^{+0.0056}_{-0.0027}$ & This Work \\
SDSS $i^\prime$ 		            & 770  & $0.0783_{-0.0037}^{+0.0036}$ & This Work \\
SDSS $z^\prime$ 		            & 869  & $0.077^{+0.005}_{-0.004}$    & \cite{hirano2016} \\
SDSS $J^\prime$ 		            & 1246 & $0.063^{+0.007}_{-0.007}$    & \cite{hirano2016} \\
SDSS $H^\prime$ 	                & 1639 & $0.073^{+0.007}_{-0.007}$    & \cite{hirano2016} \\
SDSS $K_S^\prime$	   	            & 2154 & $0.086^{+0.005}_{-0.006}$    & \cite{hirano2016} \\
\textit{Spitzer} $4.5 \unit{\mu m}$ & 4500 & $0.0795^{+0.0022}_{-0.0023}$ & \cite{chen2018} \\
\textit{Spitzer} $4.5 \unit{\mu m}$ & 4500 & $0.0760 \pm 0.0019$          & \cite{guo2018} \\
\enddata
\end{deluxetable}

To further illustrate, we overlay a simulated cloud-free solar-metallicity transmission model of K2-28b in Figure \ref{fig:transmission}. We calculated the cloud-free transmission model using the freely available \texttt{PandExo} \citep{batalha2017pandexo} code, which uses the open-source \texttt{ExoTransmit} \citep{kempton2017} code to model the transmission spectra. For this simulation we assume a hydrogen-helium dominated composition with a mean molecular weight of $\mu = 2.3 m_{\mathrm{H}}$. To calculate a transmission spectrum we need the mass of K2-28b to estimate its planetary surface gravity. As the mass of K2-28b is currently unknown, then as further discussed in Section \ref{sec:rv}, we predict the mass of K2-28b using the probabilistic mass-radius relation code \texttt{Forecaster} \citep{chen2017}. Using our best-fit radius of $R_p = 2.56_{-0.27}^{+0.26} R_\oplus$ from our joint ground and \textit{K2} fits, \texttt{Forecaster} predicts a mass of $7.81 \unit{M_\oplus}^{+5.92}_{-3.08} M_\oplus$, which we use to estimate a median planetary surface gravity of $g = 11.3 \unit{m s^{-1}}$. Furthermore, in \texttt{PandExo}, we input a planetary equilibrium temperature of $T_{\mathrm{eq}} = 500 \unit{K}$, the closest available grid value to our derived equilibrium temperature of $T_{\mathrm{eq}} = 421_{-32}^{+29} \unit{K}$ from our joint-fits, assuming a bond-albedo of 0.3 (see Table \ref{tab:k228bmcmc}). Following \cite{diamondlowe2018}, the input radius used in \texttt{Exo-Transmit/PandExo} is the planet radius without an atmosphere, and thus smaller than the radius measured by transit observations by an amount that depends on the atmospheric composition, temperature, and gravity---and changing this radius alters the amplitude of the model features as well as the overall depth of the model. As these values are uncertain for K2-28b, we calculated a number of transmission spectra in steps of $0.1 R_\oplus$ close to our best-fit value of $2.5 R_\oplus$, until we achieved a transmission spectrum that visually agreed with the ground-based measurements in Figure \ref{fig:transmission}. The non-atmosphere input radius to \texttt{PandExo} that visually agreed best with the data in Figure \ref{fig:transmission}, was $2.3 R_\oplus$. No attempt was done to fit a best-atmospheric model, or look at atmospheres dominated by clouds, given the large errorbars and the limited constraints the current observations provide. We leave it to future observations either from JWST and/or the upcoming instruments on the upcoming Extremely Large Telescopes to gather further insight into K2-28b's atmosphere.

With K2-28b's flux peaking in the NIR, the instruments on JWST provide a compelling opportunity to study K2-28b's atmosphere, as has been mentioned by many groups \citep{hirano2016,chen2018}. To quantify the feasibility to observe the transmission spectrum of K2-28b, in Figure \ref{fig:transmission} we also show the expected JWST spectrum for K2-28b for the same cloud-free model calculated using \texttt{PandExo}, using NIRISS in Single Object Slitless Spectroscopy (SOSS) mode after 5 transit observations of K2-28b binned to a resolving power of $R=50$. This simulation assumes a 1x transit length before and after the transit as an out-of-transit baseline. We calculated the transmission spectrum for NIRISS in SOSS mode, as NIRISS in SOSS mode yields the highest information content for any single JWST disperser combination \citep{batalha2017information}. \texttt{Pandexo} optimized the observation to use the \texttt{GR700XD} grism available for NIRISS. Although we will not know the actual precision of JWST until after launch and commissioning, for these calculations we follow \cite{greene2016} and assume a flat systematic noise floor of 20ppm. From Figure \ref{fig:transmission}, we see that if K2-28b has a clear atmosphere, JWST/NIRISS will have sufficient sensitivity to discern between a structured clear atmosphere, and a cloudy flat atmosphere\footnote{We decided to omit a cloudy flat-line transmission spectrum from Figure \ref{fig:transmission} for clarity.}. If K2-28b is observed to have a flat transmission spectrum, that would be consistent with the rising statistical trends of flat cloudy spectra for cold Neptune planets \citep{crossfield2017trends}. As further mentioned by \cite{chen2018}, for planets with flat transmission spectra, secondary eclipse observations offer an important and complementary window into the atmospheres of such planets, and they show that K2-28b is the only small $<3 R_\oplus$ and cool $<600 \unit{K}$ planet aside from GJ 1214b with a potentially detectable secondary eclipse with JWST.

However, due to the degeneracies between atmospheric compositions and planet surface gravities \citep{batalha2017}, to use the full potential of transmission modeling, there is a great need to measure K2-28b's mass. We further discuss the possibility of following up K2-28b with radial velocity measurements in the following subsection.

\subsection{Possibility for future RV observations}
\label{sec:rv}
Due to their brightness, both K2-28b and K2-100b are amenable for follow-up observations with precise radial velocity spectrographs to measure their masses. Doing so will give key insights into the composition of both planets, and also break important degeneracies present in transmission-modeling, as was discussed above. As mentioned further above, measuring the mass of K2-100b will yield informative insight into the masses of relatively young planets (800 Myr), being one of few transiting planets in a cluster known to orbit a bright enough planet-host for precision RV observations. Lastly, obtaining a number of high-precision radial velocity observations, can also allow us to discern if there might be more planets present orbiting these systems.

\begin{figure*}[t]
	\begin{center}
		\includegraphics[width=0.95\textwidth]{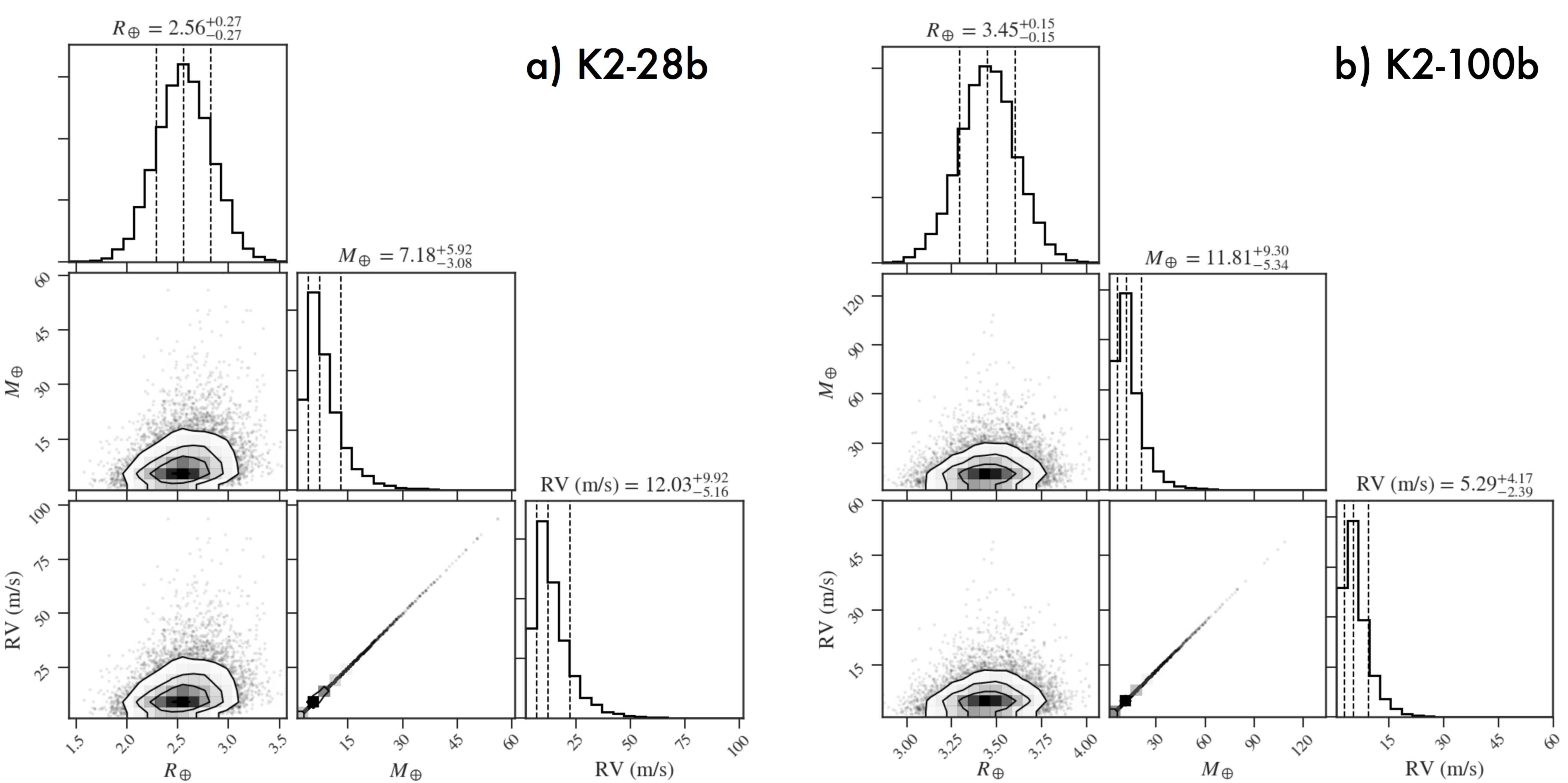}
	\end{center}
	\caption{Predicted masses and radial velocity semi-amplitudes for a) K2-28b, b) K2-100b. Masses are predicted from the best-fit radii using Forecaster \citep{chen2017}, and the expected RV semi-amplitude is calculated using Equation \ref{eq:rv}, using the predicted mass and the best-fit parameters obtained from the joint-fit models in this work as inputs.}
	\label{fig:rv_mass}
\end{figure*}

To calculate the expected RV semi-amplitude of both planets, we predicted the masses for both K2-28b and K2-100b using the Forecaster \citep{chen2017} probabilistic Mass-Radius-relation modeling code, using the best-fit radii and other orbital parameters from our MCMC fits as inputs. Figure \ref{fig:rv_mass} shows the expected mass, and RV semi-amplitude posteriors for both planets. We estimate a mass of $7.18^{+5.92}_{-3.08} M_\oplus$, and $11.81^{+9.30}_{-5.34} M_\oplus$, for K2-28b and K2-100b, respectively. For K2-28b, we note that we predict a slightly lower mass than reported in \cite{chen2018} of $8 M_\oplus$. Using these predicted masses, Forecaster classifies both planets as 100\% Neptunian, where the transition point between Terran and Neptunian composition is defined as $2.04^{+0.66}_{-0.59} M_\oplus$, and between Neptunian and Jovian as $0.414^{+0.057}_{-0.065} M_J$. This is consistent with what we would expect from the radius gap presented in \cite{fulton2017}. Using these predicted mass values, we calculated the expected RV semi-amplitude $K$, using Equation 14 from Lovis \& Fischer 2010,
\begin{equation}
K = \frac{28.4329 \unit{m s^{-1}}}{\sqrt{1-e^2}} \frac{m_2 \sin i}{M_\mathrm{Jup}} \left( \frac{m_1 + m_2}{M_\odot} \right)^{-2/3} \left( \frac{P}{1 \unit{yr}} \right)^{-1/3},
\label{eq:rv}
\end{equation}
where $e$ is the eccentricity of the planet orbit, $i$ is the best-fit inclination from the transit, $m_1$ is the mass of the star, $m_2$ is the (predicted) mass of the planet, and $P$ is the orbital period. For the masses of the star, we assumed that the masses were Gaussian distributed about the mean and the $\sigma$ given in Table \ref{tab:stellarpriors}. From Equation \ref{eq:rv}, we estimate RV semi-amplitudes of $12.03^{+9.92}_{-5.16} \unit{m/s}$ and $5.29^{+4.17}_{-2.39}\unit{m/s}$ for K2-28b, and K2-100b, respectively. Being an M-dwarf, K2-28 will be most efficiently observed with precise ultra-stabilized near-infrared radial velocity spectrographs, including e.g., CARMENES \citep{quirrenbach2012}, the stabilized Habitable-zone Planet Finder (HPF; \cite{mahadevan2012,stefansson2016}), the Infrared Doppler Instrument (IRD) for Subaru \citep{kotani2014}, or Spirou \citep{artigau2014}. Meanwhile K2-100b would be most efficiently be observed in the optical by e.g., CARMENES \citep{quirrenbach2012}, ESPRESSO \citep{pepe2014}, EXPRES \citep{jurgenson2016}, G-CLEF \citep{szentgyorgyi2012}, HARPS \citep{mayor2003}, or NEID \citep{schwab2016}, and/or other instruments in the growing worldwide network of precision radial velocity spectrographs \citep{wright2017}.

\section{Summary and Conclusion}
We present two high precision ground-based transits of the recently discovered Neptune-sized planets K2-28b and K2-100b, using the ARCTIC imager on the ARC 3.5m Telescope at APO. To achieve high precision photometry, we use the Engineered Diffuser recently commissioned on ARCTIC \citep{stefansson2017}, allowing us to maintain a broad and stable top-hat PSF throughout the night to maximize our observing efficiency. We achieve a precision of $1532 \pm 101$ppm and $358 \pm 20$ppm in 1 minute bins, and a precision of $105^{+87}_{-37}$ppm and $38^{+21}_{-11}$ppm in 30 minute bins, for K2-28b and K2-100b, respectively. These high-cadence, high-precision observations allow us to resolve the transit shape better than in the under-sampled V-shaped \textit{K2} transits, yielding improved constraints on the orbital parameters for both transits. For K2-100b, the observations presented here is the first ground-based light curve of this challenging-to-observe transit, yielding a transit depth of $819 \pm 50 \unit{ppm}$ in the SDSS $i^\prime$ band, consistent with the \textit{K2} transit depth of $689^{+6}_{-5}\unit{ppm}$ at the $2-3\sigma$ level. 

Jointly modeling our ground-based transits with data from \textit{K2}, we provide an order of magnitude improvement in the orbital ephemerides from our \textit{K2}-analysis alone, allowing us to predict the timing of transits to within 2 minutes for both planets at the start of the JWST era in 2021. For K2-28b, we compare our ephemeris updates to the recent work by \cite{chen2018} using \textit{Spitzer} observations, showing a slight disagreement in the timing of the transits in the early JWST era or at the $\sim$10min level ($3-4 \sigma$). Although modestly discrepant, this disagreement will not preclude from efficient scheduling of transits with JWST, but it highlights the importance of more than one transit follow-up observations to assess the the timing accuracy of transit ephemerides.

Our K2-28b light-curve in the SDSS i$^\prime$ band adds to the growing body of broad-band measurements of its radius. To look at the emerging picture of K2-28b's transmission spectrum, we compare available the planet radius measurements for K2-28b from this work and from broad-band transit observations from the ground \citep{hirano2016} and from \textit{Spitzer} (\cite{guo2018,chen2018}). In doing so, we show that the currently available radius measurements are not precise enough to distinguish between a structured clear atmosphere or a cloudy flat atmosphere. Using the online predictive tool \texttt{PandExo}, we demonstrate that JWST/NIRISS in SOSS mode will have enough sensitivity in a modest number of transits ($\sim$5) to discern if K2-28b has a cloudy or clear atmosphere.

We show that there is a great need to measure the masses of K2-28b and K2-100b. First, to accurately model the transmission spectra of K2-28b, its mass is essential to break degeneracies that exist between atmospheric composition and surface gravity \citep{batalha2017}. Second, K2-100b is one of few planets known to transit a host star in a cluster bright enough ($V=10.5$) to enable precision radial velocity observations. With this in mind, we use the probabilistic mass-radius relation in the \texttt{Forecaster} \citep{chen2017} package to predict that K2-28b has a mass of $7.18^{+5.92}_{-3.08} M_\oplus$ and an RV semi-amplitude of $K = 12.03^{+9.92}_{-5.16} \unit{m s^{-1}}$. Similarly, we predict that K2-100b has a mass of $11.81^{+9.30}_{-5.34} M_\oplus$ and an RV semi-amplitude of $K = 5.29^{+4.17}_{-2.39} \unit{m s^{-1}}$. These values demonstrate that both planets are within reach of current high-precision radial velocity spectrographs in the optical and/or NIR.

We conclude that diffuser-assisted photometry can play an important role in the TESS era to perform timely and precise follow-up of the expected bounty of planet candidates that TESS is expected to find.

All of the photometry and associated \texttt{Python} analysis code for this paper is captured in Jupyter notebooks freely accessible on GitHub\footnote{\url{https://github.com/gummiks/Diffuser-Assisted-K2-Followup}}.

\acknowledgments
%We thank the anonymous referee for a thoughtful reading of the manuscript, and for useful suggestions and comments. 
We gratefully acknowledge the work and assistance of Tasso Sales and Laura Weller-Brophy at RPC Photonics, without whose help this project would not have been possible. We thank Natasha Batalha for useful comments regarding exoplanet transmission spectroscopy. We thank the observing staff at Apache Point Observatory for all their help with these observations, with special thanks to Jack Dembicky, Candace Gray, William Ketzeback, Russet McMillan and Theodore Rudyk.

This work was directly seeded and supported by a Scialog grant from the Research Corporation for Science Advancement (Rescorp) to SM, LH, JW. This work was partially supported by funding from the Center for Exoplanets and Habitable Worlds. The Center for Exoplanets and Habitable Worlds is supported by the Pennsylvania State University, the Eberly College of Science, and the Pennsylvania Space Grant Consortium. GKS wishes to acknowledge support from NASA Headquarters under the NASA Earth and Space Science Fellowship Program-Grant NNX16AO28H.

These results are based on observations obtained with the Apache Point Observatory 3.5-meter telescope which is owned and operated by the Astrophysical Research Consortium. This paper includes data collected by the \textit{Kepler} telescope. The \textit{K2} data presented in this paper were obtained from the Mikulski Archive for Space Telescopes (MAST), through the Everest pipeline. Space Telescope Science Institute is operated by the Association of Universities for Research in Astronomy, Inc., under NASA contract NAS5-26555. Support for MAST for non-HST data is provided by the NASA Office of Space Science via grant NNX09AF08G and by other grants and contracts. Funding for the \textit{K2} Mission is provided by the NASA Science Mission directorate. We acknowledge support from NSF grants AST-1006676, AST-1126413, AST-1310885, AST-1517592, the NASA Astrobiology Institute (NAI; NNA09DA76A), and PSARC. This research made use of the NASA Exoplanet Archive, which is operated by the California Institute of Technology, under contract with the National Aeronautics and Space Administration under the Exoplanet Exploration Program. 

Facility: ARC 3.5m, \textit{K2}

Software:
\texttt{AstroImageJ} \citep{collins2017},
\texttt{Astroplan} \citep{morris2018},
\texttt{Astropy} \citep{astropy2013},
\texttt{Astroquery} \citep{ginsburg2016},
\texttt{batman} \citep{kreidberg2015}, 
\texttt{corner.py} \citep{dfm2016}, 
\texttt{emcee} \citep{dfm2013}, 
\texttt{Everest 2.0} \citep{luger2017}, 
\texttt{Exo-Transmit} \citep{kempton2017},
\texttt{Jupyter} \citep{jupyter2016},
\texttt{matplotlib} \citep{hunter2007},
\texttt{MC3} \citep{cubillos2017},
\texttt{numpy} \citep{vanderwalt2011},
\texttt{pandas} \citep{pandas2010},
\texttt{PandExo} \citep{batalha2017pandexo},
\texttt{pyde} \citep{pyde}

% ---------------------------
% Bibliography
% ---------------------------
\pagebreak
\bibliographystyle{yahapj}
\bibliography{references}

\newpage

\appendix
\section{Best fit parameters for K2-28b and K2-100b}
\begin{deluxetable*}{llcccc}
\tabletypesize{\small}
\tablecaption{Best fit MCMC parameters for K2-28b, including $1\sigma$ errorbars. Values are shown for our \textit{K2}-only, diffuser-assisted ground-based observations only, and joint \textit{K2} and ground-based analyses, respectively. For our joint fits, we had independent parameters for the $R_p/R_*$ values in the two different bands (\textit{K2} band, and SDSS i$^\prime$), resulting in slightly different values derived for the planet radii, transit depths, transit durations, and transit ingress/egress durations. These values are separated with a (K2), or (Ground), respectively. The eccentricity and argument of periastron were fixed at 0 for all fits.\label{tab:k228bmcmc}}
\tablehead{              \colhead{~~~Parameters} &                \colhead{Description}  &                            \colhead{K2}&                        \colhead{Ground}&                      \colhead{Joint fit (adopted)} }
\startdata
                  $T_{0}$ $(\mathrm{BJD_{TDB}})$ &                      Transit Midpoint &  $2456977.98986_{-0.00027}^{+0.00026}$ &  $2457931.89782_{-0.00039}^{+0.00039}$ &   $2457931.89780_{-0.00037}^{+0.00036}$ \\
                                      $P$ (days) &                        Orbital period &     $2.260443_{-0.000020}^{+0.000020}$ &        $2.26044_{-0.00010}^{+0.00010}$ &   $2.2604455_{-0.0000010}^{+0.0000010}$ \\
                                   $R_p/R_*$ (K2)&                          Radius ratio &           $0.0763_{-0.0027}^{+0.0056}$ &                                      --&            $0.0817_{-0.0034}^{+0.0032}$ \\
                              $R_p/R_*$ (Ground) &                          Radius ratio &                                     -- &           $0.0783_{-0.0036}^{+0.0037}$ &            $0.0790_{-0.0032}^{+0.0032}$ \\
                            $R_p (R_\oplus)$ (K2)&                         Planet radius &                 $2.42_{-0.25}^{+0.27}$ &                                     -- &                  $2.56_{-0.26}^{+0.27}$ \\
                        $R_p (R_\oplus)$ (Ground)&                         Planet radius &                                     -- &                 $2.46_{-0.26}^{+0.28}$ &                  $2.48_{-0.26}^{+0.26}$ \\
                                 $R_p (R_J)$ (K2)&                         Planet radius &              $0.216_{-0.023}^{+0.025}$ &                                     -- &               $0.229_{-0.024}^{+0.024}$ \\
                             $R_p (R_J)$ (Ground)&                         Planet radius &                                     -- &              $0.219_{-0.023}^{+0.025}$ &               $0.221_{-0.023}^{+0.024}$ \\
                                    $\delta$ (K2)&                         Transit depth &        $0.00583_{-0.00040}^{+0.00089}$ &                                      --&         $0.00667_{-0.00054}^{+0.00054}$ \\
                               $\delta$ (Ground) &                         Transit depth &                                     -- &        $0.00612_{-0.00055}^{+0.00059}$ &         $0.00625_{-0.00050}^{+0.00052}$ \\
                                         $a/R_*$ &             Normalized orbital radius &                   $16.9_{-3.8}^{+2.7}$ &                   $14.2_{-2.2}^{+2.9}$ &                    $14.2_{-1.7}^{+2.4}$ \\
                                        $a$ (AU) &                       Semi-major axis &           $0.0222_{-0.0050}^{+0.0044}$ &            $0.019_{-0.0036}^{+0.0043}$ &            $0.0191_{-0.0029}^{+0.0037}$ \\
$\rho_{\mathrm{*,transit}}$ ($\mathrm{g/cm^{3}}$)&                       Density of star &                  $17.9_{-9.6}^{+10.0}$ &                   $10.7_{-4.3}^{+8.0}$ &                    $10.7_{-3.4}^{+6.4}$ \\
                                $i$ $(^{\circ})$ &                   Transit inclination &                   $88.2_{-1.5}^{+1.3}$ &                 $87.18_{-0.97}^{+1.1}$ &                  $87.1_{-0.74}^{+0.90}$ \\
                                             $b$ &                      Impact parameter &                 $0.54_{-0.34}^{+0.22}$ &               $0.701_{-0.19}^{+0.095}$ &                 $0.72_{-0.14}^{+0.075}$ \\
                                             $e$ &                          Eccentricity &                    $0.0_{-0.0}^{+0.0}$ &                    $0.0_{-0.0}^{+0.0}$ &                     $0.0_{-0.0}^{+0.0}$ \\
                           $\omega$ ($^{\circ}$) &                Argument of periastron &                    $0.0_{-0.0}^{+0.0}$ &                    $0.0_{-0.0}^{+0.0}$ &                     $0.0_{-0.0}^{+0.0}$ \\
                           $T_{\mathrm{eq}}$ (K) &  Equilibrium temp. (assuming $a=0.3$) &                $387.0_{-28.0}^{+53.0}$ &                $422.0_{-38.0}^{+39.0}$ &                 $421.0_{-32.0}^{+29.0}$ \\
                              $S$ ($S_{\oplus}$) &                       Insolation Flux &                  $15.6_{-4.0}^{+10.0}$ &                   $21.9_{-6.9}^{+9.3}$ &                    $21.9_{-6.0}^{+6.8}$ \\
                            $T_{14}$ (days) (K2) &                      Transit duration &           $0.0398_{-0.0012}^{+0.0026}$ &                                      --&            $0.0409_{-0.0017}^{+0.0018}$ \\
                        $T_{14}$ (days) (Ground) &                      Transit duration &                                      --&           $0.0416_{-0.0019}^{+0.0027}$ &            $0.0407_{-0.0016}^{+0.0017}$ \\
                               $\tau$ (days) (K2)&               Ingress/egress duration &           $0.0039_{-0.0011}^{+0.0032}$ &                                     -- &            $0.0060_{-0.0018}^{+0.0022}$ \\
                           $\tau$ (days) (Ground)&               Ingress/egress duration &                                     -- &           $0.0056_{-0.0019}^{+0.0026}$ &            $0.0058_{-0.0018}^{+0.0021}$ \\
                  $T_{S}$ $(\mathrm{BJD_{TDB}})$ &             Time of secondary eclipse &  $2456979.12008_{-0.00026}^{+0.00025}$ &  $2457933.02804_{-0.00039}^{+0.00039}$ &   $2457933.02802_{-0.00037}^{+0.00036}$ \\
\enddata
\end{deluxetable*}

\newpage
 
\begin{deluxetable*}{llccccc}
\tabletypesize{\small}
\tablecaption{Best fit MCMC parameters for K2-100b, including $1\sigma$ errorbars. Values are shown for our \textit{K2}-only, diffuser-assisted ground-based observations only, and joint \textit{K2} and ground-based analyses, respectively. For our joint fits, we had independent parameters for the $R_p/R_*$ values in the two different bands (\textit{K2} band, and SDSS i$^\prime$), resulting in slightly different values derived for the planet radii, transit depths, transit durations, and transit ingress/egress durations. These values are separated with a (K2), or (Ground), respectively. The eccentricity and argument of periastron were fixed at 0 for all fits.\label{tab:k2100bmcmc}}
\tablehead{               \colhead{~~~Parameters} &                \colhead{Description}  &                           \colhead{K2} &                      \colhead{Ground} &                    \colhead{Joint fit (adopted)} }
\startdata
                   $T_{0}$ $(\mathrm{BJD_{TDB}})$ &                      Transit Midpoint &  $2457140.71940_{-0.00029}^{+0.00029}$ &  $2457828.69239_{-0.00056}^{+0.00074}$&  $2457828.69348_{-0.00042}^{+0.00045}$ \\
                                       $P$ (days) &                        Orbital period &     $1.673911_{-0.000012}^{+0.000011}$ &      $1.673906_{-0.00005}^{+0.00005}$ &  $1.6739024_{-0.0000011}^{+0.0000012}$ \\
                                   $R_p/R_*$ (K2) &                      Radius ratio (K2)&       $0.02625_{-0.00011}^{+0.00011}$  &                                    -- &        $0.02646_{-0.00028}^{+0.00061}$ \\
                               $R_p/R_*$ (Ground) &                  Radius ratio (Ground)&                                     -- &       $0.02862_{-0.00091}^{+0.00091}$ &        $0.02864_{-0.00092}^{+0.00094}$ \\          
                             $R_p (R_\oplus)$ (K2)&                    Planet radius (K2) &                 $3.41_{-0.14}^{+0.14}$ &                                     --&                 $3.45_{-0.15}^{+0.16}$ \\
                        $R_p (R_\oplus)$ (Ground) &                Planet radius (Ground) &                                     -- &               $3.71_{-0.19}^{+0.20}$  &                 $3.71_{-0.19}^{+0.20}$ \\
                                 $R_p (R_J)$ (K2) &                     Planet radius (K2)&              $0.304_{-0.013}^{+0.013}$ &                                    -- &              $0.308_{-0.014}^{+0.014}$ \\
                             $R_p (R_J)$ (Ground) &                 Planet radius (Ground)&                                     -- &             $0.331_{-0.017}^{+0.018}$ &              $0.331_{-0.017}^{+0.018}$ \\
                                     $\delta$ (K2)&                     Transit depth (K2)&    $0.000689_{-0.000005}^{+0.000006}$  &                                     --&       $0.000700_{-0.00002}^{+0.00003}$ \\
                                 $\delta$ (Ground)&                 Transit depth (Ground)&                                     -- &      $0.000819_{-0.00005}^{+0.00005}$ &        $0.00082_{-0.00005}^{+0.00005}$ \\
                                          $a/R_*$ &             Normalized orbital radius &              $8.181_{-0.15}^{+0.093}$  &                $7.36_{-0.67}^{+0.37}$ &                 $7.77_{-0.81}^{+0.42}$ \\
                                         $a$ (AU) &                       Semi-major axis &           $0.0452_{-0.0020}^{+0.0020}$ &          $0.0404_{-0.0036}^{+0.0030}$ &           $0.0426_{-0.0044}^{+0.0031}$ \\
$\rho_{\mathrm{*,transit}}$ ($\mathrm{g/cm^{3}}$) &                       Density of star &                 $3.7_{-0.20}^{+0.13}$  &                $2.69_{-0.67}^{+0.43}$ &                 $3.17_{-0.90}^{+0.54}$ \\
                                 $i$ $(^{\circ})$ &                   Transit inclination &                $88.98_{-0.69}^{+0.69}$ &                  $87.5_{-1.9}^{+1.7}$ &                   $87.5_{-2.0}^{+1.7}$ \\
                                              $b$ &                      Impact parameter &              $0.146_{-0.098}^{+0.094}$ &                $0.32_{-0.21}^{+0.19}$ &                 $0.34_{-0.23}^{+0.20}$ \\
                                              $e$ &                          Eccentricity &                    $0.0_{-0.0}^{+0.0}$ &                   $0.0_{-0.0}^{+0.0}$ &                    $0.0_{-0.0}^{+0.0}$ \\
                            $\omega$ ($^{\circ}$) &                Argument of periastron &                    $0.0_{-0.0}^{+0.0}$ &                   $0.0_{-0.0}^{+0.0}$ &                    $0.0_{-0.0}^{+0.0}$ \\
                            $T_{\mathrm{eq}}$ (K) &  Equilibrium temp. (assuming $a=0.3$) &               $1060.0_{-17.0}^{+17.0}$ &              $1119.0_{-34.0}^{+53.0}$ &               $1089.0_{-33.0}^{+61.0}$ \\
                               $S$ ($S_{\oplus}$) &                       Insolation Flux &                $877.0_{-55.0}^{+58.0}$ &            $1090.0_{-130.0}^{+220.0}$ &              $980.0_{-110.0}^{+240.0}$ \\
                             $T_{14}$ (days) (K2) &                  Transit duration (K2)&        $0.06627_{-0.00037}^{+0.00036}$ &                                     --&        $0.06682_{-0.00045}^{+0.00056}$ \\
                         $T_{14}$ (days) (Ground) &             Transit duration (Ground) &                                     -- &           $0.071_{-0.0014}^{+0.0013}$ &        $0.06698_{-0.00047}^{+0.00058}$ \\
                               $\tau$ (days) (K2) &          Ingress/egress duration (K2) &       $0.001739_{-0.00004}^{+0.00007}$ &                                     --&        $0.00194_{-0.00022}^{+0.00055}$ \\
                           $\tau$ (days) (Ground) &      Ingress/egress duration (Ground) &                                     -- &         $0.0022_{-0.00022}^{+0.00052}$&        $0.00210_{-0.00023}^{+0.00058}$ \\
                   $T_{S}$ $(\mathrm{BJD_{TDB}})$ &             Time of secondary eclipse &  $2457141.55635_{-0.00029}^{+0.00029}$ &  $2457829.52935_{-0.00056}^{+0.00075}$&  $2457829.53043_{-0.00042}^{+0.00045}$ \\
\enddata
\end{deluxetable*}

\end{document}